\documentclass[letter]{aa} 

\usepackage[english]{babel} 
\usepackage{natbib}
\usepackage{amssymb}
\usepackage{amsmath}
\usepackage{url}
\usepackage{txfonts}
\usepackage{rotating} 
\usepackage{xspace}
\usepackage{color}
\usepackage[switch]{lineno} 
\usepackage{txfonts}
\usepackage{caption}
\usepackage{graphicx}

\newcommand{\pmn}{PMN\,J1603--4904\xspace}
\newcommand{\pks}{PKS\,1718--649 \xspace}
\usepackage{lineno}

\begin{document}

   \title{The MHz-peaked radio spectrum of the unusual
     $\gamma$-ray source PMN\,J1603--4904}

   \author{
     C.~M\"uller\inst{\ref{affil:ru}}
     \and P.~R.~Burd\inst{\ref{affil:wuerzburg}}
     \and R.~Schulz\inst{\ref{affil:astron}}
     \and R.~Coppejans\inst{\ref{affil:ru}}
     \and H.~Falcke\inst{\ref{affil:ru}}
     \and H.~Intema\inst{\ref{affil:leiden}}
     \and M.~Kadler\inst{\ref{affil:wuerzburg}}
     \and F.~Krau\ss\inst{\ref{affil:uva}}
     \and R.~Ojha\inst{\ref{affil:nasa_gsfc},\ref{affil:umbc},\ref{affil:cua}}
     }

   \institute{
Department of Astrophysics/IMAPP, Radboud University, PO
  Box 9010, 6500 GL Nijmegen, The Netherlands \label{affil:ru}
\and Institut f\"ur
Theoretische Physik und Astrophysik, Universit\"at W\"urzburg, Am
Hubland, 97074 W\"urzburg, Germany \label{affil:wuerzburg} 
\and ASTRON, the Netherlands Institute for Radio Astronomy, Postbus 2,
7990 AA, Dwingeloo, The Netherlands\label{affil:astron} 
\and Leiden Observatory, Leiden University, Niels Bohrweg 2,
NL-2333CA, Leiden, The Netherlands\label{affil:leiden} 
\and GRAPPA \& Anton Pannekoek Institute for Astronomy, University of
Amsterdam, Science Park 904, 1098 XH Amsterdam, The Netherlands 
\label{affil:uva} 
\and NASA, Goddard Space Flight Center, Astrophysics Science
Division, Code 661, Greenbelt, MD 20771, USA \label{affil:nasa_gsfc}
\and CRESST/University of Maryland Baltimore
County, Baltimore, MD 21250, USA \label{affil:umbc} 
\and Catholic University of America, Washington, DC 20064,
USA \label{affil:cua} 
             }
   \date{\today}


 
  \abstract
   {The majority of bright extragalactic $\gamma$-ray sources are blazars. Only a few
     radio galaxies have been detected by
     \textsl{Fermi}/LAT. Recently, the GHz-peaked spectrum source
     PKS\,1718--649 was
     confirmed to be $\gamma$-ray bright, providing further evidence for the existence of a
     population of $\gamma$-ray loud, compact radio galaxies. 
     A spectral turnover in the radio spectrum in
     the MHz to GHz range is a characteristic
     feature of these objects, which are thought to be young due to
     their small linear sizes. The multiwavelength properties of the $\gamma$-ray source
     \object{PMN\,J1603--4904} suggest that it is a member of this source class.}
   {The known radio spectrum of \pmn can be
     described by a power law above 1\,GHz. Using observations from the Giant Metrewave Radio
     Telescope (GMRT) at 150, 325, and 610\,MHz, we investigate the behavior of the
     spectrum at lower frequencies to search for a low-frequency
     turnover.}
   {Data from the TIFR GMRT Sky Survey (TGSS ADR) catalog and archival
     GMRT observations were used to construct the
     first MHz to GHz spectrum of \pmn.}
   {We detect a low-frequency turnover of the spectrum and measure the
     peak position at about 490\,MHz (rest-frame), which,
     using the known relation
     of peak
     frequency and linear size,
     translates into a maximum linear source size of $\sim$1.4\,kpc.}
   {The detection of the MHz peak indicates that \pmn is part
     of this population of
     radio galaxies with turnover frequencies in the MHz to GHz regime. Therefore it can be considered the second confirmed
     object of this kind detected in $\gamma$-rays. Establishing this $\gamma$-ray
     source class will help to investigate the $\gamma$-ray production
     sites and to test broadband emission models.
}


   \maketitle
%

\section{Introduction}\label{sec:intro}
Most sources detected at $\gamma$-ray energies by the \textsl{Fermi}
Large Area Telescope (LAT) are active galactic nuclei (AGN), in
particular blazars, which are extragalactic jets observed at small
angles to the line of sight \citep{3fgl}. Only a few, nearby non-blazar
sources are detected \citep[see also,
e.g.,][]{Abdo2010_misaligned,3fgl}. In contrast to blazars, jets in
radio galaxies or misaligned sources are seen at larger
inclination angles, hence, they are less or not at all
relativistically beamed. A typical broadband spectrum 
can be explained by inverse Compton or hadronic processes in the
jet \citep[see, e.g.,][and references
therein]{Massaro2016_Fermireview}.
The
location and mechanism of the high-energy emission are
still subject to much debate. Possible origins include
core, parsec-scale jets, and the kiloparsec-scale lobes. In
nearby sources \citep[Centaurus~A and
Fornax~A;][]{Abdo2010_CenAlobes,Ackermann2016_FornaxA} the $\gamma$-ray
emission from the lobes can be directly imaged with the \textsl{Fermi}/LAT.

The smaller versions of evolved radio galaxies are of a few
kpc or less in size
\citep[e.g.,][]{Readhead1996b,Readhead1996a,Odea1998,Kunert-Bajraszewska2010}. Since their
morphologies are reminiscent of the evolved double sources, they
are called compact symmetric objects'
($\lesssim$\,1\,kpc, CSOs) or medium-sized symmetric objects'
($\lesssim$\,1-15\,kpc, MSOs).
Kinematic
measurements of their compact lobes suggest that
these sources are younger than full-sized radio galaxies \citep[e.g.,][]{Owsianik1998,An2012}.  An alternative
explanation of the smaller extensions is the scenario of frustrated jets,
i.e., the source is confined owing to the interaction with a dense,
surrounding medium \citep{Bicknell1997,Carvalho1998}. 
Thus, these sources play an important role in the study of AGN evolution and the
interaction of AGN jets with the ambient medium.

Typically, CSOs show a turnover or
cutoff in their radio flux-density spectrum around 1\,GHz, where the spectrum
changes from flat or steep at higher frequencies to an inverted
spectrum. The peak of MSOs is typically lower, in the upper
megahertz range. The convex spectral shape of their radio spectrum
is one of the defining properties of these small radio sources.
Based on the peak frequency $\nu_\mathrm{peak}$, one refers to MHz-peaked spectrum
(MPS) or compact steep spectrum (CSS) sources
for $\nu_\mathrm{peak}<1$\,GHz and GHz-peaked sources (GPS) for
$\nu_\mathrm{peak}>1$\,GHz \citep{Odea1998}.

The reason for this turnover could be either due to
synchrotron self-absorption \citep[SSA; e.g.,][]{Snellen2000} by
relativistic electrons from the emitting source, or
free-free absorption due to an external dense medium \citep[FFA; e.g.,][]{Bicknell1997}. 
Well-sampled
radio spectra are necessary to identify the underlying absorption
mechanism \citep[e.g.,][]{Callingham2015}. 

Several studies found that the linear
size $LS$ of the radio morphology anticorrelates with the rest-frame peak
frequency of the radio spectrum \citep{Fanti1995,Odea1997,Odea1998},
\begin{equation}\label{eq:Odea}
\log\nu_\mathrm{peak}  \approx  -0.21(\pm 0.05) - 0.65(\pm0.05)\log LS
,\end{equation}
with LS in kpc and $\nu_\mathrm{peak}$ in GHz.
This relation indicates that the mechanism causing the spectral
turnover is related to the source size and is explained well in the SSA scenario.
Therefore, a typical GPS radio galaxy has an extent of less than
1\,kpc. A typical MPS or CSS is up
to 20\,kpc in size according to this relation.

Evolved radio galaxies have been established as
$\gamma$-ray loud objects \citep[e.g.,][]{Abdo2010_misaligned}.
Theoretical broadband emission models also
\citep[][]{Kino2009,Kino2011,Kino2013,Stawarz2008,Ostorero2010}
predict $\gamma$-ray emission from compact, likely young, radio
galaxies. Because of the
interaction of the evolving radio source and the interstellar medium,
high-energy emission can be produced through the non-thermal inverse
Compton process or thermal bremsstrahlung. The interaction of
particles in the lobes with the ambient medium is expected to dominate
the emission, while the contribution of the jet is smaller because of
Doppler deboosting \citep{Stawarz2008}.  \citet{Kino2009} predict
thermal bremsstrahlung $\gamma$-ray emission in the initial phase of
expansion at a detectable level for \textsl{Fermi}/LAT observations of
nearby young radio galaxies.

Two CSS sources, 3FGL J1330.5+3023 (3C 286) and 3FGL
J0824.9+3916 (4C +39.23B), were reported in the latest \textsl{Fermi}/LAT AGN catalog \citep[3LAC,][]{3lac}.
However, GPS sources are still
not confirmed as a $\gamma$-ray bright source class. 
Only recently, \citet{Migliori2016}
reported of the first detection of an established CSO/GPS,
\object{PKS\,1718--649}. This result follows the discussions
of a few $\gamma$-ray CSO-candidates detected by
\textsl{Fermi}/LAT \citep[see also][for a summary]{DAmmando2016}:
4C+55.1 \citep[][]{McConville2011},
PKS\,1413+135 \citep[][]{Gugliucci2005}, 2234+282 \citep{An2016}, and
PMN\,J1603--4904 \citep{Mueller2014a}.

An extensive multiwavelength study of \pmn
\citep{Mueller2014a,Mueller2015a}, conducted within the
framework of the TANAMI program \citep{Ojha2010a,Kadler2015},
suggested it was not a blazar. Very Long Baseline
Interferometry (VLBI) observations indicate a CSO-like morphology. The
kinematic study on timescales of $\sim$15\,months
shows no significant motion of the eastern and western
component. Lower resolution radio observations with the ATCA ($>$1\,GHz)
indicate extended, unresolved emission at sub-arcsecond scales.
X-ray observations with \textsl{XMM-Newton} and
\textsl{Suzaku} measure an emission line in the X-ray
spectrum on top of an absorbed power-law
emission. Together with the optical-spectroscopic redshift of $z=0.23$ from X-shooter observations
\citep{Goldoni2016} this X-ray feature could be interpreted as emission from a highly ionized
plasma, but the lack of a neutral iron line is puzzling.  The \textsl{Fermi}/LAT counterpart 3FGL\,1603.9$-$4903 (with
95\%-confidence semimajor axis of
$\sim$0.8\,arcmin)\footnote{A false association of the $\gamma$-ray source
  with PMN J1603–4904 is still possible, however, as pointed out in
  \citet{Mueller2015a} a more exotic explanation for the $\gamma$-ray
  emission origin would be required.} shows
a hard $\gamma$-ray spectrum but with a spectral break above 50\,GeV
and only mild variability \citep{1fhl,3fgl,2fhl}.  These broadband
properties led to the conclusion that \pmn is either a very peculiar
blazar or a compact, possibly young, radio galaxy.  
Opposite motion of the radio lobes and a turnover in the radio spectrum
would hence be expected and would further confirm the CSO nature.

Here we present a detailed analysis of the radio spectrum of \pmn, extending
the existing information down to 150\,MHz.


\section{Observations and data reduction}\label{sec:obs}
The analysis of the radio spectrum presented in \citet{Mueller2014a} shows
that the \pmn spectrum above $\gtrsim1$\,GHz follows a power law with
a spectral index (defined as $S_\nu \sim
  \nu^{+\alpha}$) of $\alpha\sim-0.35$ (see Sect.~\ref{sec:results}). The lowest ATCA frequency is
$\sim$1\,GHz, hence, besides one archival spectral point at
843\,GHz \citep{Murphy2007_MGPS}, no spectral information in the MHz
regime was available.

With the recent release of the catalog from the TIFR GMRT Sky Survey
\citep[TGSS ADR;][]{TGSS} from the Giant Metrewave Radio Telescope
\citep[GMRT;][]{Swarup1991}, the first low-frequency information on
\pmn became available. The TGSS catalog contains 150\,MHz all-sky survey data
obtained between 2010 and 2012, covering the sky north of $-53^\circ$
declination with an approximate resolution of $\sim 25'' \times 25''/
\cos(\mathrm{DEC-19^\circ} )$ for DEC$<$$19^\circ$, otherwise $25''$ circular.
The unresolved catalog source J160350.7$-$490405 can be associated
with \pmn. It has a flux density
of $702.3\pm71.6\,\mathrm{mJy}$ and is modeled with a single Gaussian component of
$\sim25.3'' \times 68.5''$.

\begin{figure}
\includegraphics[width=0.5\textwidth]{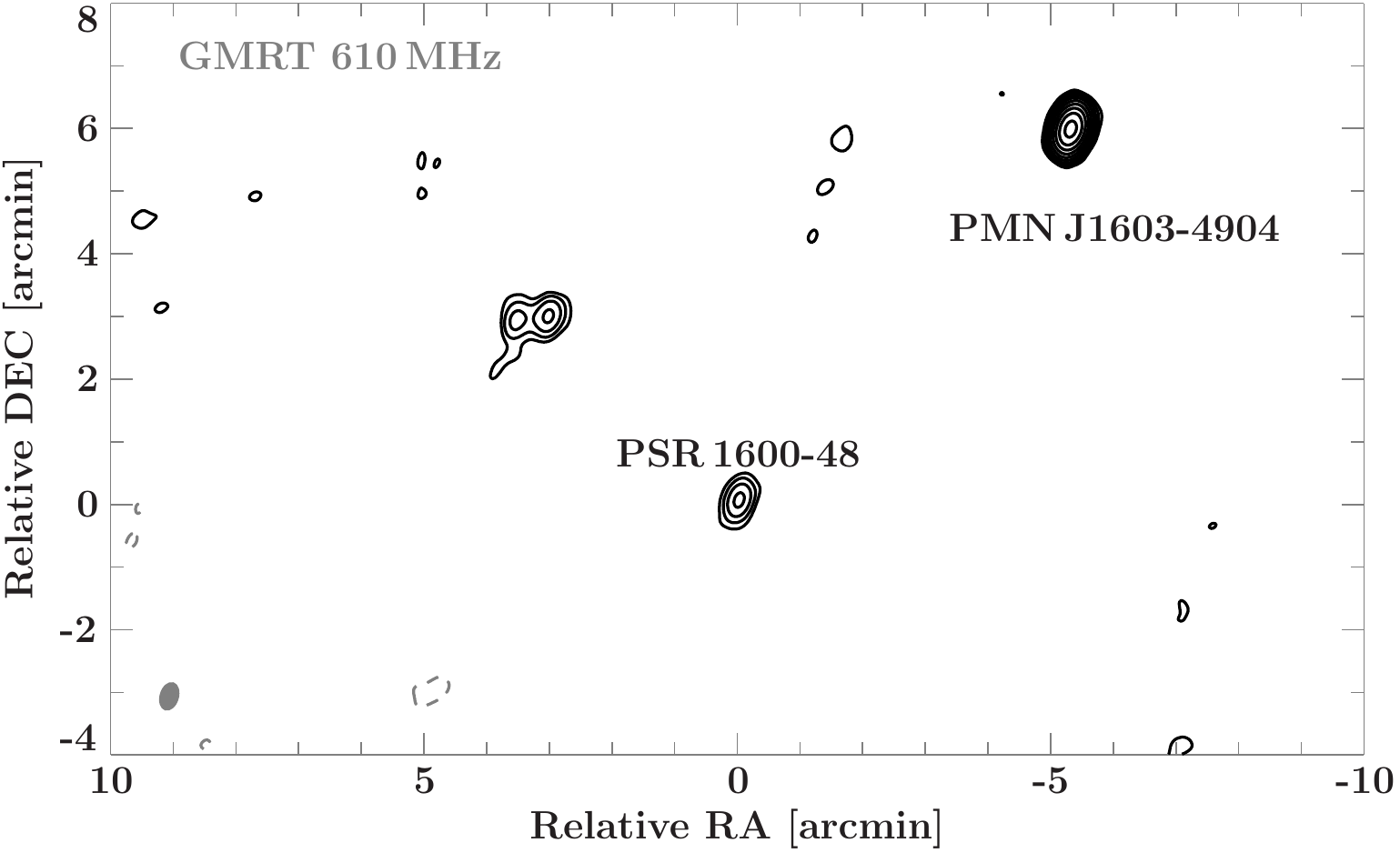}
 \caption{Radio map at 610\,MHz observed with the GMRT. The
  x- and y-axis give relative angular distances to the pulsar
  PSR\,1600$-$48 in right ascension and declination, respectively. An unclassified
   double source and \pmn, the brightest source in that region, are
   clearly detected. The contours
   indicate the flux density level, scaled logarithmically 
   with the lowest level set to the 3$\sigma$ noise level. The
   restoring beam is shown as a gray ellipse in the lower left corner.}
\label{fig:gmrt}
\end{figure}

Following this result, we analyzed archival GMRT data of \pmn at
325\,GHz (project code: 25\_038, observation date: 2014-01-31) and
610\,GHz (25\_038, obervation date: 2014-03-23). 
These observations were pointed at the pulsar
PSR\,1600$-$48. The large field of view
of the GMRT observations ($\sim 43$\,arcmin at 610\,MHz) also allowed
us to detect \pmn which has a distance from the pulsar of about $\sim$8\,arcmin.
The GMRT data were calibrated using the Astronomical Imaging
Processing System
\citep[AIPS;][]{Greisen2003_AIPS}. Since GMRT data are strongly influenced by
radio frequency interference (RFI), we used one channel that is not corrupted to perform the
amplitude calibration. Using the flux calibrator 3C\,286 we calibrated
the amplitudes of \pmn for the given channel and applied a bandpass
calibration afterward (using AIPS task \texttt{BPASS}). Note that
\pmn is not resolved, therefore we did not calibrate the phases. With the AIPS
task \texttt{FLGIT} we flagged RFI noise. 
Afterward we channel averaged
and exported the data to DIFMAP \citep{Shepherd1997}, where we used an
iterative self-calibration procedure to image the field. We used Gaussian
model fits to obtain the flux densities of the individual sources. 
We are only interested in the central part of the map (inner $\sim10$\,arcmin) and not in
the detailed structure of the sources, hence, bandwidth smearing and
non-coplanar baseline effects (w-projection) are
not an issue. 
GMRT data are prone to flux scale errors due to sky temperature
differences between the flux calibrator and target field \citep[e.g.,][]{Marcote2015}. Following \citet{TGSS}, the
radio images at 325 and 610\,MHz were corrected for this effect by a
scaling factor depending on frequency and sky position. For
\pmn these correction factors are  $3.56\pm0.2$ for 325\,MHz and $1.58\pm0.2$ for
610\,MHz. In order to account for systematic errors we add a 10\%
uncertainty \citep{Chandra2004}. A summary of the available MHz data of \pmn can be found in
Table~\ref{table1}.

Figure~\ref{fig:gmrt} shows the resulting 610\,MHz image of the field, clearly
showing the detection of the pulsar (phase center), a double source to
the northeast (TGSS-IDs:
J160444.6$-$490703 and J160444.7$-$490703) and
\pmn, which is unresolved at both frequencies.

\begin{table}
\caption[]{\em{MHz flux densities of \pmn.}} \label{table1}
\centering
\begin{tabular}{cccc}
\hline\hline
$\nu_\mathrm{obs}$& $S_\mathrm{\nu, measured}$ &
                                                            $S_\mathrm{\nu, corrected}$& Reference\\
 $[$MHz$]$&  [Jy] & [Jy]& \\

\hline
150 & --&$0.702\pm0.072$  & TGSS \\
300 & $0.61\pm0.07$& $2.18\pm0.23$ & this work \\
600 & $1.21\pm0.10$& $1.91\pm0.25$  & this work \\
843 & $1.54\pm0.5$ & -- & MGPS\\
\hline
\end{tabular}
\end{table}


\section{The MHz to GHz spectrum of \pmn}\label{sec:results}
\begin{figure}
\includegraphics[width=0.5\textwidth]{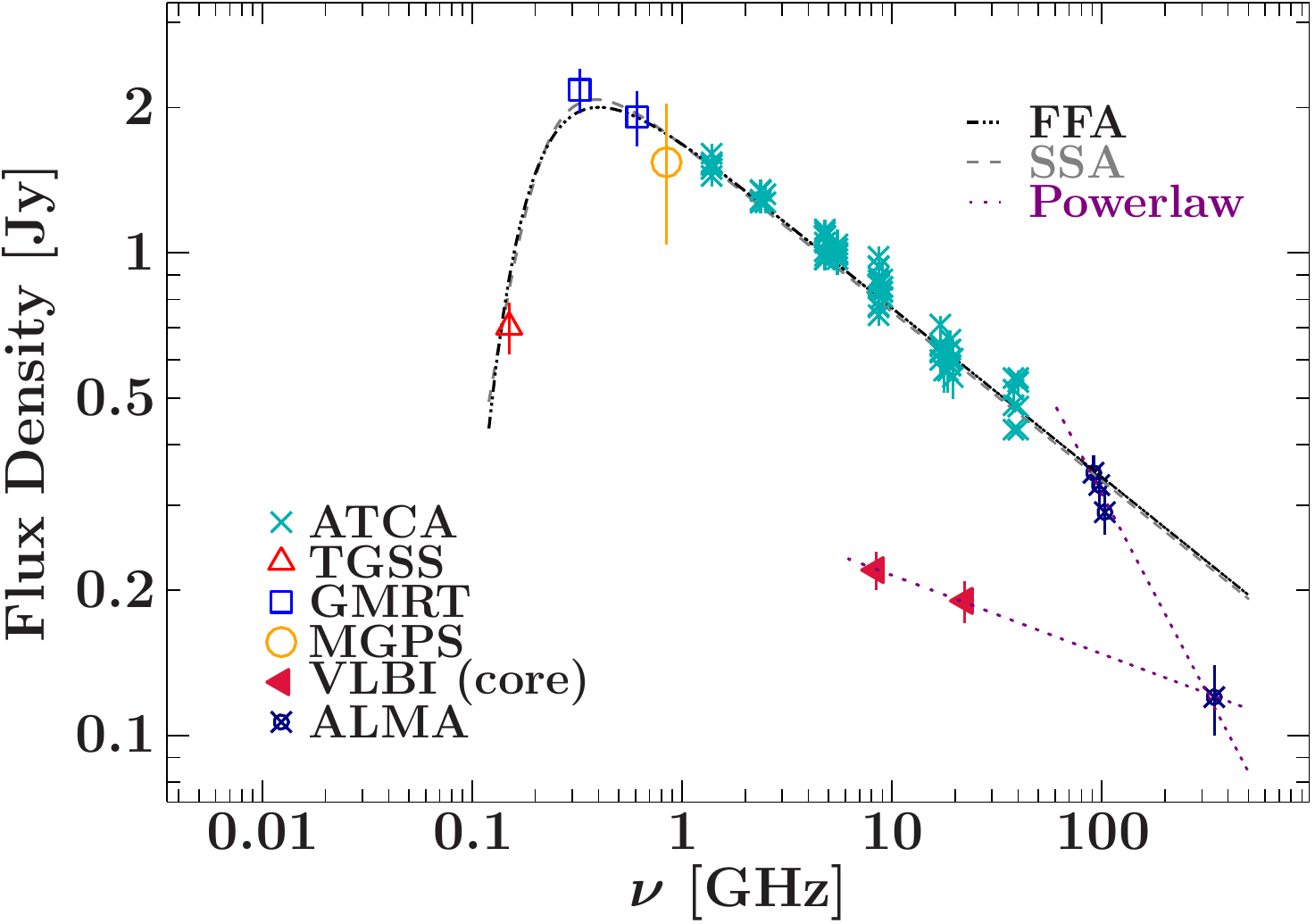}
 \caption{Radio spectrum of \pmn including archival data points
   from the ATCA, VLBI \citep{Mueller2014a}, MGPS \citep{Murphy2007_MGPS}, the ALMA
   calibrator survey, TGSS \citep{TGSS}, and GMRT (this work). The
   low-resolution data (i.e., except VLBI core fluxes) can be
 well described by a FFA (black) or SSA (gray) spectrum with a spectral
 turnover at $\sim$400\,MHz (observed frame). The
 highest measurement at 343.5\,GHz by ALMA could either be due to a
 spectral break around 100\,GHz or emission of the inner jet
 indicated by the power-law spectrum defined by the VLBI core
 (see Sect.~\ref{sec:results}).}
\label{fig:spec}
\end{figure}

The previous analysis of the GHz radio data in \citet{Mueller2014a} revealed no variability
below $\sim$20\,GHz since 2000, hence, we expect that \pmn exhibits no major flux
change at sub-GHz frequencies over this time period. With this
assumption we construct a non-simultaneous
MHz to GHz spectrum of \pmn. Figure~\ref{fig:spec} shows the resulting
radio spectrum including all available archival data points, i.e., 
the data presented in \citet{Mueller2014a}, measurements from the
ALMA calibrator
database\footnote{\url{https://almascience.eso.org/alma-data/calibrator-catalogue}}, 
and the new MHz measurements from GMRT.

The spectrum clearly shows a turnover at $\sim 400$\,MHz (observed frame). 
Such a turnover in the radio regime of an AGN spectrum can be either
due to SSA or FFA. We test both models, using the SSA model
\begin{equation}\label{eq:ssa}
S_\mathrm{\nu,SSA} = a_\mathrm{SSA} \nu^{+2.5} [1-\exp(-\tau_\mathrm{SSA}\nu^{\alpha-2.5})]
\end{equation}
and the FFA model
\begin{equation}\label{eq:ffa}
S_\mathrm{\nu,FFA} = a_\mathrm{FFA} \nu^\alpha \exp(-\tau_\mathrm{FFA}\nu^{-2.1})
\end{equation}
with the spectral index $\alpha$ and the SSA or FFA absorption
coefficient $\tau_\mathrm{SSA/FFA}$, respectively \citep[following the definition by][]{Kameno2003}. 
We use $\alpha=-0.35$ from a fit to the ATCA and ALMA data above
$>$1\,GHz. We fit all
data, which are all from unresolved images, except the core fluxes from VLBI.

Both models can describe the data equally well (see
Fig.~\ref{fig:spec} and
Table~\ref{table:fits}). The $\chi^2_\mathrm{red}$ values are comparable
to results by \citet{Callingham2015} who discussed FFA and SSA models for the GPS source \object{PKS\,B0008-421}.
The authors stress that more densely sampled data below the turnover would be required 
to distinguish between the absorption mechanisms. 
The spectral coverage allows us to constrain the spectral turnover of
\pmn at $\sim 400$\,MHz (observed frame),
but not to check more sophisticated models. 
Using the anticorrelation between rest-frame peak frequency (i.e.,
$\sim490$\,MHz at $z=0.23$) and size (see
Eq.~\ref{eq:Odea}) from
\citet[][]{Odea1997} we can estimate the maximum size of \pmn to be
$\sim 1.4$\,kpc. This is close to the canonical limit of 1\,kpc for CSOs
\citep{Odea1998} and well within the limits for MSOs.

At $\gtrsim$100\,GHz, the ALMA measurement at 343.5\,GHz (Band~7)
could indicate a break in the synchrotron spectrum. A power-law fit to
the ALMA data only yields $\alpha_\mathrm{ALMA}\sim -0.8$. Such a
spectral break could be explained by radiation losses. 
The
VLBI analysis shows that about 20\% of the extended emission detected
by the ATCA is resolved out \citep{Mueller2014a}. A power-law fit to
the VLBI core flux densities at 8 and 22\,GHz and the ALMA Band~7
measurement yields a spectral index of $\alpha_\mathrm{core}\sim
-0.16$. The Band~7 measurement basically presents the extrapolation of
the core spectrum.  
Taking the known source geometry into account, it is more plausible that the
jet or core is dominating the overall emission at these high frequencies,
while at lower energies the emission of the large-scale structure,
beyond VLBI scales, can be detected.

\begin{table}
\caption{\em{Spectral fit parameters.}} \label{table:fits}
\centering
\begin{tabular}{lcccc}
\hline\hline
Model &  $a$ & $\tau$ &
                                                  $\nu_\mathrm{turnover}$$^a$
  [GHz] & $\chi^2_\mathrm{red}$\\
\hline
FFA&  1.7$\pm$0.2 & 0.02$\pm$0.01& 0.40$\pm$0.04&4.4\\
SSA &  98$\pm$24 &  0.02$\pm$0.01 & 0.39$\pm$0.03&4.5\\
\hline
\end{tabular}
\\
\footnotesize{$^a$observed-frame peak or turnover frequency of the spectrum
}
\end{table}


\section{Conclusions}\label{sec:conclusions}

We have presented the first MHz to GHz spectrum for the $\gamma$-ray
loud extragalactic jet source \pmn. The radio spectrum clearly shows a rest-frame
spectral turnover at $\sim 490$\,MHz. Using the established 
anticorrelation of peak frequency and linear size, we determine the
linear size of \pmn to be $\sim 1.4$\,kpc. 
With VLBI, only the inner $\sim$40\,pc of the small radio
source are detected \citep{Mueller2014a}. 
The spectral index of the optically thin emission is comparably flat
for GPS or CSS sources \citep{Odea1998,Fanti2001}, and is likely due to
the bright compact emission.
The spectral shape can be explained by the superposition of
different components. Both, the SSA and FFA models can describe
the overall spectrum. The turnover can hence be attributed either to
self-absorption or to external absorption by a dense medium.

The sparse spectral coverage below 1\,GHz does not allow the testing
of more
complex models or distinguishing between SSA and FFA. More sensitive
observations with southern telescopes like the upgraded GMRT \citep{RaoBandari2013} or the Murchinson Widefield
Array \citep[MWA;][]{MWA}, covering a wider frequency
range in the MHz regime would be required.  Higher resolution
observations with sufficient sensitivity could allow us to image the extended
emission. Currently, only ALMA is capable of addressing
these scales at this low declination.

Because of its multiwavelength properties, \pmn has been discussed as a
possible $\gamma$-ray loud young radio galaxy. With the
observations reported here, this classification is supported by the
detection of an MPS-like radio spectrum. \pmn adds to the class of so far very rare
extragalactic jets that show a spectral turnover in the MHz to GHz range, and are
detected at $\gamma$-rays.
As it is the second confirmed object of this type, following
\pks \citep{Migliori2016}, there are some noteworthy
differences. First, \pks is a nearby galaxy (z=0.014) in contrast to \pmn at
z=0.23. Furthermore, \pmn has a hard GeV spectrum with a photon index 
$\Gamma\sim2$ \citep[][]{1fgl,2fgl,3fgl}, while for \pks a
$\Gamma\sim2.9$ was found \citep{Migliori2016}, that is
more comparable with the spectrum of the Cen~A lobes
\citep{Abdo2010_CenAlobes}. Hence, as also discussed by
\citet{Migliori2016}, this could indicate significant differences in the
intrinsic physical mechanism producing $\gamma$-rays. This 
is also suggested by the VLBI structure of both sources: while \pks
has an irregular lobe-dominated morphology \citep{Tingay2003,Ojha2010a}, \pmn
is a more symmetric, core dominated source.

Future \textsl{Fermi}/LAT observations using the Pass8 analysis \citep{Atwood2013_Pass8} will have
the required sensitivity to detect further GPS and CSS sources at
$\gamma$-rays. By establishing this source class and its defining
properties one will be able to test theoretical emission models
\citep[e.g.,][]{Kino2007,Kino2009,Stawarz2008}.
For \pmn, modeling of the broadband SED will help us to understand the
high-energy emission mechanism. The high Compton dominance \citep[see
Fig.~7 in][]{Mueller2014a} and the flat $\gamma$-ray spectrum challenges current
models. 


\begin{acknowledgements}
We thank the anonymous referee for the prompt review and
T.~Beuchert for helpful comments that improved the manuscript.
We thank the staff of the GMRT that made these observations
possible. GMRT is run by the National Centre for Radio Astrophysics of
the Tata Institute of Fundamental Research. 
This paper makes use of ALMA calibrator database:
\url{https://almascience.eso.org/alma-data/calibrator-catalogue}. 
ALMA is a partnership of ESO
(representing its member states), NSF (USA) and NINS (Japan), together
with NRC (Canada), NSC and ASIAA (Taiwan), and KASI (Republic of
Korea), in cooperation with the Republic of Chile. The Joint ALMA
Observatory is operated by ESO, AUI/NRAO and NAOJ.
C.M. and H.F. acknowledge support from the ERC Synergy Grant ``BlackHoleCam –
Imaging the Event Horizon of Black Holes'' (Grant
610058). F.K. acknowledges funding from the European Union’s Horizon
2020 research and innovation program under grant agreement No
653477. R.S. acknowledges support from the ERC under the European
Union's Seventh Framework Programme (FP/2007-2013)/ERC Advanced Grant
RADIOLIFE-320745.
This research was funded in part by NASA through
Fermi Guest Investigator grant NNH13ZDA001N-FERMI.
\end{acknowledgements}

\bibliographystyle{aa}
\bibliography{mnemonic,aaabbrv,pmn_gps}

\end{document}